\documentclass[12 pt, amsfonts,amsmath, amssymb,color]{article}

\usepackage{amsmath,amssymb,amsfonts,latexsym,float,graphics,epsfig}
\usepackage{subfig}
\usepackage{verbatim}
\usepackage{relsize}
\usepackage{hyperref}

\def\be{\begin{equation}}
\def\ee{\end{equation}}
\def\bea{\begin{eqnarray}}
\def\eea{\end{eqnarray}}

\begin{document}
%<<<<<<<<<<< enumeration of eqns section wise>>>>>>>>>>>>>>>>>>>

\renewcommand\theequation{\arabic{section}.\arabic{equation}}
\catcode`@=11 \@addtoreset{equation}{section}
%<<<<<<<<<<<<<<<<<<<<<<<<<<<<<<<<<>>>>>>>>>>>>>>>>>>>>>>>>>>>>>>>>>
\newtheorem{axiom}{Definition}[section]
\newtheorem{theorem}{Theorem}[section]
\newtheorem{axiom2}{Example}[section]
\newtheorem{lem}{Lemma}[section]
\newtheorem{prop}{Proposition}[section]
\newtheorem{cor}{Corollary}[section]

\newcommand{\ben}{\begin{equation*}}
\newcommand{\een}{\end{equation*}}
\title{\bf Thermodynamic curvature of AdS black holes with dark energy}
\author{
\bf Aditya Singh\footnote{E-mail: as52@iitbbs.ac.in}, \hspace{0.5mm} Aritra Ghosh\footnote{E-mail: ag34@iitbbs.ac.in} \hspace{0.5mm} and Chandrasekhar Bhamidipati\footnote{E-mail: chandrasekhar@iitbbs.ac.in} \\
~~~~~\\
 School of Basic Sciences, Indian Institute of Technology Bhubaneswar,\\   Jatni, Khurda, Odisha, 752050, India\\
}

\date{ }

\maketitle

\begin{abstract}
    In this paper, we study the effect of dark energy on the extended thermodynamic structure and interacting microstructures of black holes in AdS, through an analysis of thermodynamic geometry. Considering various limiting cases of the novel equation of state obtained in charged rotating black holes with quintessence, and taking enthalpy $H$ as the key potential in the extended phase space, we scrutinize the behavior of the Ruppeiner curvature scalar $R$ in the entropy-pressure $(S,P)$-plane (or equivalently in the temperature-volume ($T,V$)-plane). Analysis of $R$ empirically reveals that dark energy parameterized by $\alpha$, significantly alters the dominant interactions of neutral, charged and slowly rotating black hole microstructures. In the Schwarzschild-AdS case: black holes smaller than a certain size continue to have attractive interactions whereas larger black holes are completely dominated by repulsive interactions which arise to due dark energy. For charged or rotating AdS black holes with quintessence, $R$ can change sign at multiple points depending upon the relation between $\alpha$ and charge $q$ or angular momentum $J$. In particular, above a threshold value of $\alpha$,  $R$ is never negative at all, suggesting heuristically that the repulsive interactions due to quintessence are long ranged as opposed to the previously known short ranged repulsion in charged AdS black holes. A mean field interaction potential is proposed whose extrema effectively capture the points where the curvature $R$ changes sign.
\end{abstract}

\section{Introduction}
Cosmological data points towards an accelerated expansion of our universe due to the presence of a large negative pressure leading to the conjectured existence of an anti-gravitational force, namely dark energy~\cite{Riess,Perlmutter,Sahni:1999gb}. Despite the prevalence of several theoretical models in efforts to explain astronomical observations, the key traits of dark energy such as its source and nature still remain mysterious. Some of the models for dark energy employ a cosmological constant $\Lambda$ or a quintessence field. $\Lambda$ is a very small constant whereas quintessence energy is quite inhomogeneous. Quintessence can be thought of as a fluid that fills the spacetime everywhere \cite{Caldwell,Chen,Azreg} and is modeled as a time dependent scalar field \(\phi(t)\) evolving towards its potential minimum. It is associated with pressure, \(P_q = \dot{\phi}^2/2 - U(\phi)\) and energy density, \(\rho_q = \dot{\phi}^2/2 + U(\phi)\) where \(U(\phi)\) is the potential function. Often, it is useful to define a quintessential state parameter as, \(\omega_q = P_q/\rho_q\) and it is clear that in general \(\omega_q \in (-1,1)\) with the extreme limits corresponding respectively to \(\dot{\phi} \rightarrow 0\) and \(U(\phi) \rightarrow 0\). It is generally believed that the late time evolution is because of the cosmological constant or quintessence field permeating through the universe. Therefore, if the quintessence fluid encompasses the whole universe, it is expected to also surround the black holes changing their spacetime geometry both near the horizon as well as at the asymptotic cosmological horizon. In this context, a new spherically symmetric quintessence black hole solutions  were found in \((3+1)\)-dimensions by Kiselev~\cite{Kiselev} and the solution in AdS space time is well studied in the literature from thermodynamics point of view~\cite{quint1}-\cite{quint2}. Here, we should caution the reader about the usage of the terminology of quintessence in Kiselev's black hole metric which may be different from the notion used in cosmology modeled by a scalar field as discussed above. Few reasons follow. First, Kiselev's metric may not be a solution of field equations derived from a gravitational action coupled to specified matter Lagrangian (such as representing quintessence field). Instead, the form of energy-momentum tensor in Kiselev's approach is a specific choice with free parameters, picked in a way to give features of quintessence and inserted in to the field equations. Furthermore, the stress-energy tensor used in the Kiselev's metric is anisotropic and does not represent a perfect fluid. Therefore, our usage of the terminology of quintessence in Kiselev black hole spacetimes considered in this paper is within the limitations of these conditions (please see~\cite{Visser} for more clarity on these issues). Despite these limitations, Kiselev's quintessence black hole metric is an interesting toy model, as being spacetimes which can be explored, with some appealing mathematical as well as physical features. For special limiting values of the equation of state parameter (to be discussed later) the pure quintessence geometry without mass or charge/rotation parameters can model spacetimes with asymptotic geometry which is a softer version of de Sitter or anti de Sitter spacetimes~\cite{Gibbons} and is of interest. We also see in this paper that the additional contribution due to quintessence helps in developing our understanding of the behavior of black hole microstructues in a broader setting with additional parameters.

\smallskip

An interesting aspect of black hole thermodynamics ensues while generalizing the notion of ADM mass from asymptotically flat to (A)dS spacetimes, namely the requirement of a dynamical cosmological constant. A variable $\Lambda$ plays a central role in extending Smarr's formula from flat to AdS spacetimes, typically giving rise to the notion of bulk pressure $P$ and a novel concept of thermodynamic volume $V$~\cite{Kastor,Cvetic:2010jb}. This novel set up where the first law is augmented by a $V dP$ term coming from a dynamical $\Lambda$ is known as extended black hole thermodynamics \cite{Hawking:1982dh}-\cite{Dolan:2011xt}. In this approach, the ADM mass $M$ of the spacetime is identified with its enthalpy (rather than internal energy $U$) as $M = H=U + PV$, with the new first law being written as,
\begin{equation} \label{FLbulk}
dH = T dS + V dP \, ,
\end{equation}
where $V= (\partial H/\partial P)_S$ is the thermodynamic volume conjugate to $P$. For charged black holes, there turns out to be an exact identification of its phase transitions~\cite{Hawking:1982dh,Chamblin:1999tk,Chamblin:1999hg} (including the matching of critical exponents) with the well known liquid-gas transitions of a van der Waals fluid (vdW)~\cite{Kubiznak:2012wp}, putting them in the same universality class \cite{Kubiznak:2012wp}-\cite{Majhi}. Thermodynamic quantities of charged rotating black holes in AdS in extended phase space were found by Gunasekaran et al. in~\cite{45}. Results were extended to Gauss-Bonnet-AdS black holes by Cai et al. and Wei et al. in~\cite{fa2,fa}. All the above set ups have also been generalized to black holes in AdS with quintessence in interesting works. For instance, effects of quintessence on the \(PV\)-critical behavior of charged AdS black holes was worked out in~\cite{Li} showing that the critical quantities are modified, but the small/large black hole behavior remains the same. Further, the effect of dark energy on the efficiency of black hole heat engines has been studied in~\cite{Liu:2017baz} showing that quintessence can improve the efficiency, especially at the critical point. Maxwell's equal area law was used by Guo in~\cite{48} to show that phase transitions of AdS black holes with quintessence can still be mapped to those of the liquid-gas system. More recently, it has been shown that quintessence alters the high temperature phase structure of black holes undergoing the Hawking-Page transition \cite{HP}. In fact, a novel black hole solution of charged rotating black holes in AdS in the presence of quintessence, generalizing the Kiselev's metric was obtained by Xu and Wang \cite{26a}. For this new solution, the thermodynamics and phase transitions in the extended thermodynamic space were discussed recently giving rise to a novel equation of state \cite{Sharif:2020hid}. Our aim is to take these studies \cite{26a,Sharif:2020hid} forward to study the geometry of the spaces of thermodynamic equilibrium states for these black hole solutions and to get additional insights into the possible microstructures of these black holes.

\smallskip

The fact that a black hole horizon can be assigned the notion of a temperature indicates towards a microscopic description of black holes where microstructures share the degrees of freedom of the Bekenstein-Hawking entropy \cite{Bekenstein:1973ur}-\cite{Bardeen:1973gs}. These microscopic structures have associated with them, thermal degrees of freedom and respect equipartition theorems (see for example \cite{Padmanabhan}). With this picture in mind, it is useful to think of the microstructures of the black hole in parallel with the molecules that constitute a fluid system in standard thermodynamics and the study of thermodynamic geometry proves to be of outmost importance in its usefulness in probing the nature of interactions between the microstructures \cite{Cai}-\cite{BTZ}. To the best of our knowledge, Ruppeiner geometry for black holes was first used in~\cite{Cai}, to gain a statistical understanding of the underlying degrees of freedom, with extensions to Reissner-Nordstr\"{o}m, Kerr and Reissner-Nordstr\"{o}m-AdS black holes with internal energy and electric potential (or angular velocity for Kerr black hole case) as the fluctuation variables~\cite{Shen}. For example, based on empirical analysis, it is now known that charged AdS black holes are associated with both attraction and repulsion dominated regions as shown in remarkable works \cite{Wei2019a,Wei2019b}. Extending to theories with higher derivative terms, including those with a Gauss-Bonnet term in the action, the attraction-repulsion dominated regions are determined by their electric charge (and Gauss-Bonnet coupling, \(\alpha_{GB}\) in \(d \geq 6\)) whereas, their neutral counterparts are dominated by attraction \cite{Xu,Wei,Ghosh,Zhou:2020vzf}. Charged BTZ black holes (see \cite{BTZ} and references therein) on the other hand, are associated with purely repulsive behavior. More recently, Ruppeiner geometry has given intriguing results for charged black holes with spherical \cite{Wei:2020kra} and hyperbolic horizons \cite{Yerra:2020tzg}, with deep insights in Hawking-Page transition and Renormalization group (RG) flows, respectively. The effect of quintessence (along with a cloud of strings) on phase transitions of charged black holes in AdS was pursued in \cite{Moumni,quint2} recently. In general, thermodynamic geometry \cite{Ruppeiner} and its generalizations have been the subject of various exhaustive studies involving several classes of thermodynamic systems including quantum gases, magnetic systems \cite{Ruppeiner2}-\cite{Janyszek1990} (see also \cite{RuppeinerRMP,interactions,Rupppppp}) and most notably black holes. The generic form of the Ruppeiner metric is defined as the negative Hessian of the entropy,
\begin{equation}\label{Ruppeiner}
  dl_R^2 = - \frac{\partial^2 S}{\partial x^i \partial x^j} dx^i dx^j ,
\end{equation} where \(x^i\) and \(x^j\) are independent thermodynamic fluctuation coordinates \(i,j \in \{1,2,...,n\}\). The entropy $S=S(x^i)$ is a function of other thermodynamical variables and the coordinates $x^i$ are conserved charges of the black holes such as $M, Q, J$ and also span over the new variable $P, V$ in extended thermodynamics. For example, in the case of Schwarzschild-AdS black holes one has \(S=S(H,P)\) where symbols have their usual meaning from standard thermodynamics. This may be inverted to obtain \(H = H(S,P)\) so that in the thermodynamic limit, one has the familiar first law\footnote{For a rotating black hole
%in extended thermodynamics where \(S\) and \(V\) are independent,
one can write the first law as: \(dH = TdS + VdP + \Phi dQ + \Omega dJ\) (in the usual notation) where \(H = U + PV\) is the enthalpy of the spacetime pointing towards the fact that the entropy is a function of the parameters, i.e. \(S = S(H,Q,J,P)\). However, for simplicity we shall completely fix \(J\) and \(Q\) as mere parameters and consider fluctuations of two thermodynamic variables only.}: \(dH = TdS + VdP\). In fact, one can write another related Weinhold metric on the space of thermodynamic equilibrium states in the enthalpy $H=M$ representation\footnote{The Weinhold metric is originally defined in the energy representation \cite{Weinhold}, taking $U$ as the key potential. In standard black hole thermodynamics $U$ is identified with mass $M$ of the black hole to evaluate the metric on the space of thermodynamic equilibrium states. However, in extended thermodynamics $M$ is identified with enthalpy $H$ and hence it is appropriate to call it an enthalpy representation.} as
\begin{equation}\label{Weinhold}
  dl_W^2 = \frac{\partial^2 H}{\partial y^i \partial y^j} dy^i dy^j
\end{equation} where \(\{y^i\}\) are independent thermodynamic variables, such as $(S,P)$. It is not difficult to show that \(dl_R^2 = dl_W^2/T\) and therefore the metrics differ only by a conformal factor (please see appendix for a concise derivation).

\smallskip

\textbf{Motivation and plan:} In this work, thermodynamic geometry is pursued in extended thermodynamics where enthalpy in eqn (\ref{FLbulk}) is identified as the correct thermodynamic potential~\cite{Kastor} for the proper identification of microstructure interactions with quintessence, keeping the vdW analogy in mind. Using enthalpy as the key potential, the effect of dark energy on microstructures using thermodynamic geometry is appropriately studied in $ (T,V)$ and $(S,P)$-planes, following references~\cite{Wei2019a}-\cite{BTZ} and~\cite{Xu,Ghosh}. We consider various limiting cases arising from the novel equation of state obtained recently in \cite{Sharif:2020hid} and use Ruppeiner's thermodynamic geometry to show that the inclusion of quintessence leads to novel repulsion dominated regions, modifying the behavior of microstructures of neutral, charged and rotating black holes in AdS. Furthermore, we introduce a phenomenological model of a mean field interaction potential for the analysis in the extended thermodynamics setting where a direct analogy with fluid systems is available. It is shown that the extrema of the interaction potential correspond to the points where the Ruppeiner curvature is zero, essentially capturing the location where the type of dynamic interactions shifts from attractive to repulsive or vice-versa.\\ ~\\The organization of the paper is as follows. In sub-section-(\ref{knbhd}) below, we start by writing down the thermodynamic relations for general charged rotating black holes in AdS surrounded by quintessence field.  Section-(\ref{Ruppeiner}) contains the basic tools of Ruppeiner geometry required to find the thermodynamic curvature. The metric on thermodynamic space in the novel \((T,V)\)- and \((S,P)\)-planes is obtained and a general formula for Ruppeiner curvature is reviewed. In section-(\ref{main}) we obtain exact results for thermodynamic curvature and study the points where it vanishes for the special cases of Reissner-Nordstr\"{o}m and Schwarzschild black holes in AdS, surrounded by quintessence. The case of slowly rotating black holes with quintessence is studied numerically in following subsection. In section-(\ref{mean}), we propose a general scheme for understanding the behavior of microstructures of AdS black holes from a mean field perspective, which in particular, captures the points where the Ruppeiner curvature vanishes. The case of charged AdS black holes (without quintessence) is discussed first in subsection-(\ref{meanRN}) and then in subsection-(\ref{quint}), we discuss the effect of quintessence on the microstructures of charged and neutral black holes in AdS using the interaction potential. Finally, we end with remarks in section-(\ref{remarks}). Some introductory discussions on thermodynamic metric structures, particularly in the context of extended black hole thermodynamics can be found in the appendix.

\subsection{Thermodynamics of AdS Kerr-Newman black holes surrounded by dark energy} \label{knbhd}
Making use of Newman-Penrose formalism \cite{25f}, the solution for a Kerr-Newman-AdS black hole generalizing the Kiselev's model metric with quintessence, was obtained by Xu and Wang \cite{26a}, with the line-element in Boyer-Lindquist coordinates given as,
\begin{equation}\label{1}
ds^{2}=-\frac{\chi}{\Omega}\bigg[dt-\frac{a\sin^2\theta}{k}
d\phi\bigg]^2 +\frac{\Omega}{\chi}dr^{2}+\frac{\Omega
}{\tilde{P}}d\theta^{2}+\frac{\tilde{P}\sin^{2}\theta}{\Omega
}\bigg[a dt-\frac{(r^2+a^2)}{k}d\phi\bigg]^2\, .
\end{equation}
Here,
\begin{eqnarray}\label{2}
\chi&=&(r^2+a^2)(1+\frac{r^2}{l^2})-2mr+q^2-\alpha r^{1-3\omega_q}, \\
\label{3} \Omega&=&r^2+a^2\cos^2\theta, \quad k=1-\frac{a^2}{l^2},
\quad \tilde{P}=1-\frac{a^2}{l^2}\cos^2\theta ,
\end{eqnarray}
where $a$ is the rotation parameter and $q^2=q_e^2+q_m^2$ with $q_e$ and $q_m$ being the electric and magnetic charges respectively. Further, $m$ is the black hole mass and $l=\sqrt{-3/\Lambda}$ is the AdS radius.
In general, the state parameter $\omega_q$ is bounded as $-1<\omega_q<-\frac{1}{3}$ and $\alpha$
is the parameter standing for the intensity of the
quintessential field surrounding a black hole, obeying the inequality \cite{26a},
\begin{equation}
\alpha\leq\frac{2}{1-3\omega_q}8^{\omega_q}.
\end{equation}
Quintessence has been introduced as an alternative to the cosmological constant scenario to account for the current acceleration of the universe. This new dark energy component allows values of the equation of state parameter \(\omega_q \geq -1\). The case of a cosmological constant corresponds to the value \(\omega_q=-1\). Kiselev~\cite{Kiselev} noted that the value of \(\omega_q = -2/3\) is favorable on symmetry grounds (the general metric is symmetric as function of radial coordinate $r$), in addition to having an asymptotic behavior reminiscent of de Sitter spacetimes with interesting horizon structure. Moreover, this case also satisfies the dominant energy condition \cite{Hawking,Gibbons}. The analysis of \(PV\)-criticality and other studies in charged AdS black holes surrounded by quintessence~\cite{Li,Liu:2017baz}  showed that another appealing feature is that analytic results could be obtained for this value, where as for other values, one has to resort to numerical methods. We thus focus on the case \(\omega_q = -2/3\) henceforth, which allows an analytic study of the mean field potential and its connection to zeroes of Ruppeiner scalar in following sections. %This relation holds true until the existence of cosmological horizon determined by quintessential dark energy.
Note that for $\alpha=0$, the line element (\ref{1}) reduces to the
Kerr-Newman-AdS solution while the Kerr-AdS solution can be obtained
by further setting $q=0$. We explore these and other limits in subsequent sections.

\smallskip

\begin{figure}[h]
%	 \begin{wrapfigure}{l}{0.3\textwidth}
	\begin{center}
		\centering
		\includegraphics[width=4.6in]{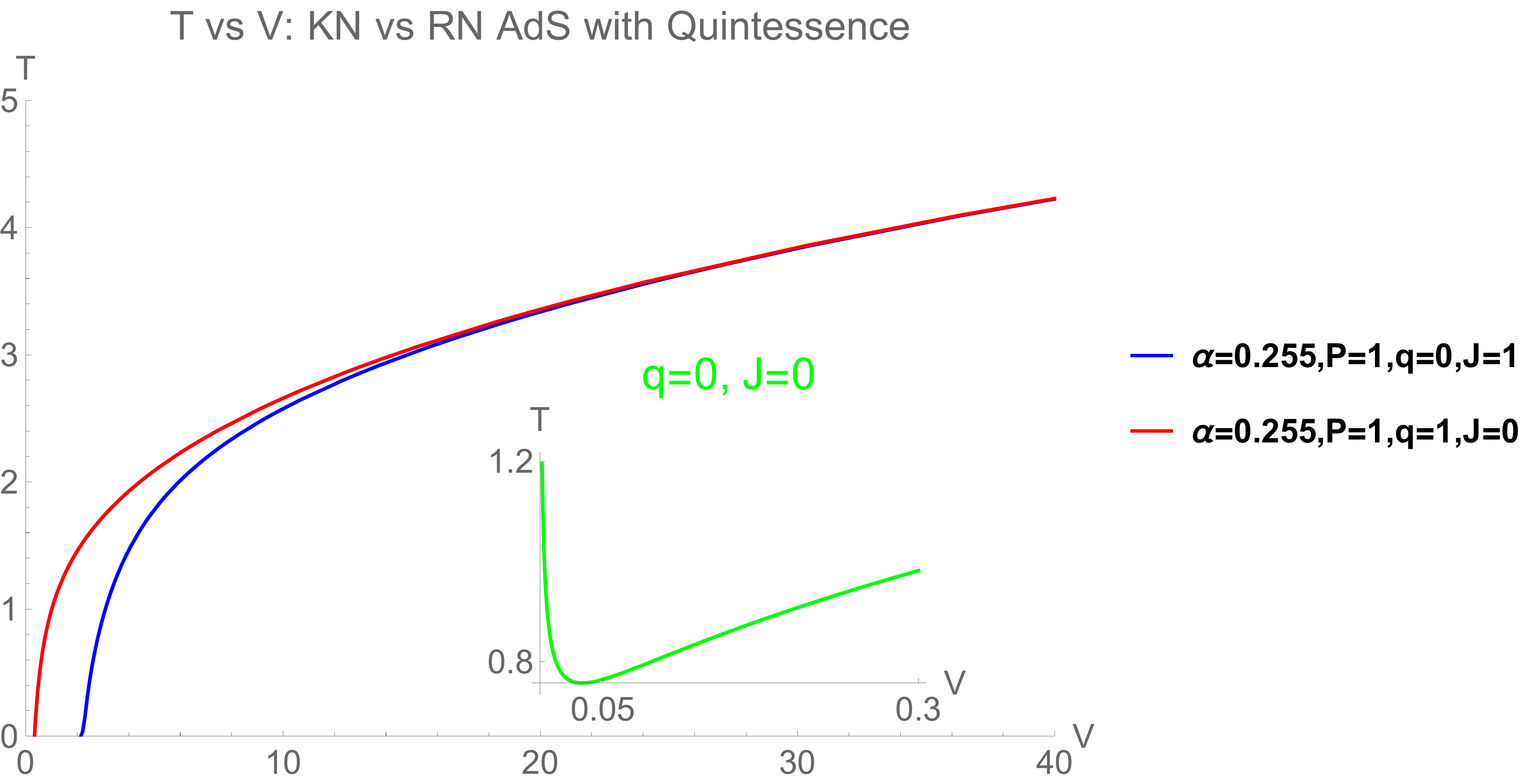}  		
		\caption{Plot of the Hawking temperature vs thermodynamic volume.}
        \label{t_vs_v_alpha}	
	\end{center}
%	\end{wrapfigure}
%	\end{wrapfigure}
\end{figure}

Useful expressions for mass, entropy and temperature are given respectively as,
\begin{equation}\label{9}
m=\frac{1}{2 r_+}(r_+ \left(\frac{r_+
\left(a^2+l^2+r_+^2\right)}{l^2}-\alpha r_+^{2}\right)+a^2+q^2),
\end{equation}
\begin{equation}\label{10}
S=\frac{A}{4}=\frac{\pi  \left(a^2+r_{+}^2\right)}{k},
\end{equation}
\begin{equation}\label{11}
T =\frac{1}{4 \pi \left(a^2+r_+^2\right)}\bigg(2 \big(\frac{r_+
\left(a^2+l^2+2 r_+^2\right)}{l^2}-m\big)-3\alpha  r_+^{2}\bigg).
\end{equation}
In extended black hole thermodynamics, the cosmological constant is identified with
pressure as,
\begin{equation}\label{P}
  P = - \frac{\Lambda}{8\pi} = \frac{3}{8\pi l^2} \, ,
\end{equation}
with its thermodynamic conjugate volume obtained as,
\begin{equation}\label{18}
V=\left(\frac{\partial M}{\partial P}\right)_{S,Q,J,\alpha}=\frac{2
\pi  \left(a^2 l^2 \left(q^2-\alpha r_+^3\right)+\left(a^2+r_+^2\right) \left(a^2 l^2-a^2 r_+^2+2 l^2
r_+^2\right)\right)}{3 k^2 l^2 r_+}.
\end{equation}
Here mass $M$, angular momentum $J$ and charge $Q$, are related to
parameters $m$, $a$ and $q$ as follows,
\begin{eqnarray}
M=\frac{m}{k^2}, \quad J=\frac{ma}{k^2}, \quad Q=\frac{q}{k}.
\end{eqnarray}
The physical parameters $M$, $J$ and $Q$ now satisfy the Smarr-Gibbs-Duhem relation \cite{Sharif:2020hid},
\begin{eqnarray}\label{20}
M&=&2(TS - PV + \Pi J ) + Q\Phi+\alpha  \Psi \left(-\frac{2
a^2}{a^2+r_+^2}-1\right),
\end{eqnarray}
where the electric potential $\Phi$ and $\Psi$
(which is conjugate to $\alpha$) \cite{mm} are given as,
\begin{eqnarray}
\Phi&=&\frac{q r_+} {a^2+r_+^2},  \quad \Psi=-\frac{r_+^{2}}{2
k}.
\end{eqnarray}
Let us note that when $\alpha=0$, all the above quantities go back to the relations for
charged rotating AdS black holes \cite{45}. The Hawking temperature in terms of pressure can be written as \cite{Sharif:2020hid},
\begin{equation}\label{24}
T=\frac{2 r_+ \left(8 \pi  P \left(a^2+2 r_+^2\right)+3\right)-r_+
\left(8 \pi  a^2 P+3\right)-\frac{3 \left(a^2+q^2\right)}{r_+}-8 \pi
P r_+^3 - 6 \alpha r_+^2}{12 \pi  \left(a^2+r_+^2\right)}.
\end{equation}
The plot of temperature vs volume is shown in figure-(\ref{t_vs_v_alpha}).
The above equation needs to be inverted to obtain the equation of state of the general Kerr-Newman AdS black holes. However, as noted in~\cite{45}, it is in general not possible to obtain an exact expression and hence one resorts to the slowly rotating charged black hole case, where the equation of state is \cite{Sharif:2020hid},
\begin{eqnarray} \label{eosqj} \nonumber
P&=&\frac{T}{2 r_+}+\frac{Q^2}{8 \pi r_+^4}-\frac{1}{8 \pi
r_+^2}+\frac{ \alpha }{4 \pi r_+}\\\nonumber&+&\frac{3
J^2 \left(Q^2 r_+^2-2 Q^4-8 \pi Q^2 r_+^3 T - 4 \alpha Q^2 r_+^3 + 8 \pi r_+^5 T - 2 \alpha  r_+^5 + 4 r_+^4\right)}{8 \pi r_+^6
\left(2 Q^2+2 \pi r_+^3 T-\frac{1}{2} \alpha r_+^3 + r_+^2\right){}^2}\\  \label{25}&+&O(J^4),
\end{eqnarray}
which is an expansion in powers of $J$ and terms of order $O(J^4)$ are neglected in further analysis.
In the following sections, various limiting cases of the above equation of state are considered leading to neutral, charged and (slowly) rotating black holes in AdS with quintessence. For each of the cases, the behavior of  thermodynamic curvature is studied and for the former two cases, a mean field potential description is also obtained.

\section{Ruppeiner geometry and microscopic interactions}\label{Ruppeiner}
To begin with, consider a thermodynamic system with some \(n\) independent thermodynamic variables which can fluctuate about their mean values. A prototypical example with \(n=2\) would be a hydrostatic system with a fixed volume described by the grand canonical ensemble. The external bath controls the temperature (\(\beta\)) and chemical potential (\(\mu\)) of the system whereas their respective thermodynamic conjugates, namely energy\footnote{Typically, temperature is conjugate to the entropy as is clear from the first law, \(dE = TdS + \) other terms. However, inverting this and writing as, \(dS = \beta dE + ....\), leads to the notion of \(\beta\) being conjugate to energy.} and number of particles are allowed to have fluctuations about their mean values. If the system is at equilibrium, which we shall assume throughout this work, the first derivatives of the entropy shall vanish identically which means that expanding \(S\) about its equilibrium value \(S_0\) to include effects of fluctuations one can write up to the lowest non-trivial order,
\begin{equation}\label{Sexpansion}
  S \approx S_0 + \frac{\partial^2 S}{\partial x^i \partial x^j} dx^i dx^j,
\end{equation} where \(\{x^i\}\) are the fluctuation coordinates the entropy is a function of and \(i = 1,2,....,n\). It is to be understood that the derivatives are evaluated at thermodynamic equilibrium. Recalling Boltzmann's definition of entropy as a measure of the probability as \(S = \ln \Omega\), one can invert this writing \(\Omega = e^S\) and then the probability of finding the system with the independent variables between \(x^i\) and \(x^i + dx^i\) for all \(i\) is then given by,
\begin{equation}\label{Pdx}
  P dx^1 dx^2 .... dx^n = C e^{-dl_R^2} dx^1 dx^2 .... dx^n,
\end{equation} where we have used eqn (\ref{Sexpansion}) and have written,
\begin{equation}\label{R}
  dl_R^2 = - \frac{\partial^2 S}{\partial x^i \partial x^j} dx^i dx^j,
\end{equation}
Clearly, eqn (\ref{R}) defines a length element on the space of independent thermodynamic variables describing the system. However, such a length has great physical meaning. Indeed, from eqn (\ref{Pdx}) it is clear that for any two points on the space of independent thermodynamic variables, the probability of them being related by a fluctuation more, the closer they are! The reader is referred to \cite{RuppeinerRMP,interactions,Rupppppp} for a much more detailed account on the role of the Ruppeiner metric in thermodynamic fluctuation theory. Note that taking any arbitrary thermodynamic potential, \(\phi = \phi(z^i)\) with new coordinates $z^i$, one can define a metric as follows,
\begin{equation}\label{newM}
  dl^2 = -\frac{\partial^2 \phi}{\partial z^i \partial z^j} dz^i dz^j.
\end{equation} This would still in some sense define a length between different thermodynamic equilibrium states. However, in general it may not have a good connection with thermodynamic fluctuation theory and may not result in curvature scalar with nice interpretation (like the one provided by Ruppeiner scalar $R$) as the measure of interactions of the system. The metric in eqn (\ref{newM}) may be meaningful if it is obtained from
eqn (\ref{R}) by an appropriate coordinate transformation. More details are given in the appendix.
%However, a new metric may be obtained from
%eqn. (\ref{newM}) by a proper coordinate transformation.  An example is the case with \(\phi = U\) which is the Weinhold metric where the fluctuation coordinates would typically be \(S\) and \(V\) (other variables can also arise in addition if there are other work terms in the first law, \(dU = TdS - PdV\)). In this case, the metric in eqn. (\ref{newM}) may be obtained from the one in eqn. (\ref{R}), by a proper coordinate transformation. (as mentioned at the end of the appendix).
%However, for static black holes in extended thermodynamic framework, where \(S\) and \(V\) are not independent variables, it would be appropriate to define such a metric as the Hessian of the enthalpy wherein fluctuation variables are rather \(S\) and \(P\) which are independent. However, although such a choice would define a length, note that the length so defined from the negative Hessian of the entropy is related to the probability of fluctuations as we just saw. Therefore, the thermodynamic length due to the Weinhold (or a Weinhold-like) metric does not have a nice physical interpretation as compared to that defined from the Ruppeiner metric.
%As remarked in the introduction however, the lengths can be conformally related meaning that the length defined from the Weinhold metric shall be physically meaningful in the context of fluctuation theory if it is scaled by an appropriate conformal factor which in this case turns out to be \(T^{-1}\).

\subsection{Line element and Ruppeiner curvature}

To proceed with the study of thermodynamic geometry of the black holes in AdS with quintessence background, we note that even though the Ruppeiner metric [eqn (\ref{Ruppeiner})] is initially defined as the Hessian of the entropy, one can calculate equivalent forms of the metric using other potentials such as the enthalpy, \(H=H(S,P)\) in eqn (\ref{Hquint}), where the fluctuation coordinates are simply \(S\) and \(P\). We shall, in this work, study Ruppeiner geometry on the \((T,V)\)- and \((S,P)\)-planes, i.e. using both enthalpy and Helmholtz free energy representations\footnote{As also pointed out in \cite{Xu}, an internal energy representation, i.e. the \((S,V)\)-plane is not suitable for many black hole systems where the fluctuation coordinates \(S\) and \(V\) are not independent.}. The line element on the \((S,P)\)-plane can be calculated without much difficulty and reads \cite{Xu, Ghosh} (see appendix for derivation),
\begin{equation}\label{RuppeinerSP}
  dl_R^2 = \frac{1}{C_P}dS^2 + \frac{2}{T}\bigg(\frac{\partial T}{\partial P}\bigg)_SdSdP - \frac{V}{TB_S}dP^2 ,
\end{equation} where \(B_S = -V(\partial P/\partial V)_S\) is the adiabatic bulk modulus which diverges for non-rotating black holes (in \(d \geq 4\)) obtained by setting \(J=0\) wherein, the entropy and volume are not independent. Similarly, the Ruppeiner line element on the \((T,V)\)-plane is given by \cite{Xu, Ghosh},
\begin{equation}\label{RuppeinerTV}
  dl_R^2 = \frac{1}{T}\bigg(\frac{\partial P}{\partial V}\bigg)_TdV^2 + \frac{2}{T}\bigg(\frac{\partial P}{\partial T}\bigg)_VdTdV + \frac{C_V}{T^2}dT^2 ,
\end{equation}
where \(C_V\) is the specific heat at constant volume which is zero for black holes where the geometric volume coincides with the thermodynamic volume. In four or more dimensions, this is true for all static (non-rotating) black holes. Since we consider the slowly rotating black hole approximation, \(C_V\) shall still be taken to be zero. A derivation of these line elements is presented at the end in the appendix.

\smallskip

Now that it is understood that the Ruppeiner metric defines a physically meaningful distance on the spaces of thermodynamic equilibrium states, let us go further and analyze the meaning of the corresponding Ricci scalar. For a two dimensional metric, i.e. with two fluctuation coordinates as we shall be using in this work, the components of the metric can be written down as a \(2 \times 2\) matrix, \(g_{ij}\) with \(i = 1,2\). The Ricci scalar of the geometry described by the Ruppeiner metric can be calculated by the rules of Riemannian geometry. This Ricci scalar shall be called the Ruppeiner curvature and contains the physical information of the microscopic interactions in a thermodynamic system.
For such a metric, the Ricci scalar can be written down as (see for example \cite{Rupppppp}),
\begin{equation}\label{Genericcurvature}
  R = - \frac{1}{\sqrt{g}}\bigg[\frac{\partial}{\partial x^1}\bigg(\frac{g_{12}}{g_{11}\sqrt{g}}\frac{\partial g_{11}}{\partial x^2} - \frac{1}{\sqrt{g}}\frac{\partial g_{22}}{\partial x^1}\bigg) + \frac{\partial}{\partial x^2}\bigg(\frac{2}{\sqrt{g}}\frac{\partial g_{12}}{\partial x^1} - \frac{1}{\sqrt{g}}\frac{\partial  g_{11}}{\partial x^2} - \frac{g_{12}}{g_{11}\sqrt{g}}\frac{\partial g_{11}}{\partial x^1}\bigg) \bigg],
\end{equation} where \(g=g_{11}g_{22}-g_{12}g_{21}\) is the determinant of the \(2 \times 2\) matrix \(g_{ij}\).
The Ruppeiner curvature calculated for any set of fluctuation coordinates for the classical ideal gas identically comes out to be zero whereas that for the vdW fluid comes out to be negative over the entire physical region\footnote{See for example the recent work \cite{Wei2019b} where the Ruppeiner curvature of the vdW fluid in the novel \((T,V)\)-plane has been presented.} At the first instance, it appears that the Ruppeiner curvature may indicate whether interactions are present in the system. If instead, one looks at ideal quantum gases like the ideal Bose or the ideal Fermi gas, by virtue of being `ideal' such gases are expected to be weakly interacting like the classical ideal gas meaning that the only interactions between particles can be the collisions. However, computation of the Ruppeiner curvature for ideal Bose and Fermi gases give opposite signs \cite{Janyszek1990}. In the sign convention adopted in this work (opposite to that of \cite{Janyszek1990}), the ideal Bose gas is associated with a Ruppeiner curvature which always carries a negative sign whereas, that for the Fermi gas carries a positive sign. It is known that fermions typically have repulsive interactions of quantum mechanical origin between them as a direct consequence of the exclusion principle which leads to the familiar notion of a Fermi pressure. The Ruppeiner curvature being negative definite for an attractive vdW fluid and being positive definite for a repulsive quantum gas of fermions strongly suggests its relationship with the nature of interactions between the underlying degrees of freedom of the system. Further, an ideal gas of bosons has with it associated a negative definite thermodynamic curvature which is also consistent with the fact that bosons tend to come closer to each other, an interaction which is yet again of purely quantum mechanical origin. The Ruppeiner curvature therefore, not only probes the vdW type interactions, but also interactions arising out of the quantum mechanical nature of the underlying degrees of freedom and as such, it is a perfect macroscopic probe which may be used to understand how the microscopic constituents of a given thermodynamic system interact \cite{RuppeinerRMP,interactions,Rupppppp} (see also \cite{RupMay}). This deserves special attention for the case of black holes where a microscopic theory is not well understood and hence, Ruppeiner geometry is expected to provide us with early microscopic insights.

\section{Effect of dark energy on thermodynamic geometry} \label{main}
%%%%%%%%%%%%%%%%%%%%%%%%%%%%%%
From the form of the equation of state given in eqn (\ref{eosqj}), it is helpful to first discuss the thermodynamic geometry of cases $J=0,Q=0$ and $J=0, Q \neq 0$, corresponding to neutral and charged black holes in AdS, respectively surrounded by quintessence, where an exact analysis is possible. In these case, it is also possible to explicitly see the form of mean field potential (section-(\ref{mean})). The other case of interest, i.e., $Q=0, J \neq 0$ is discussed in subsection-(\ref{slowly}), separately as the expressions for thermodynamic curvature are quite involved and the behavior can only be studied numerically.

\subsection{Schwarzschild and Reissner-Nordstr\"{o}m black holes in AdS}

\begin{figure}[h]
%	 \begin{wrapfigure}{l}{0.3\textwidth}
	\begin{center}
		\centering
		\includegraphics[width=4.6in]{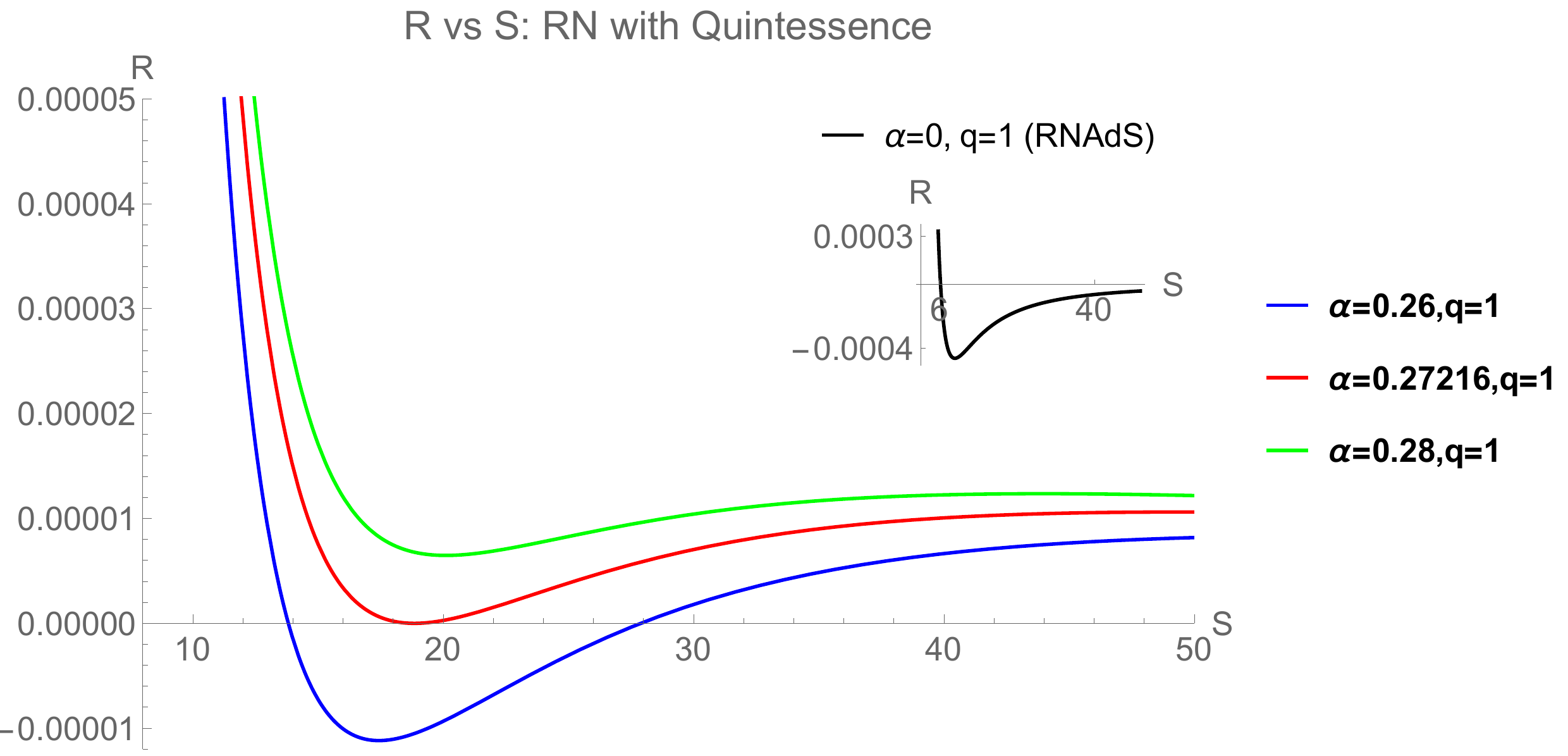}  		
		\caption{Ruppeiner curvature on the \((S,P)\)-plane as a function of \(S\) for fixed \(P\).}
        \label{RSplot}	
	\end{center}
%	\end{wrapfigure}
\end{figure}

\begin{figure}[h]
%	 \begin{wrapfigure}{l}{0.3\textwidth}
	\begin{center}
		\centering
		\includegraphics[width=4.0in]{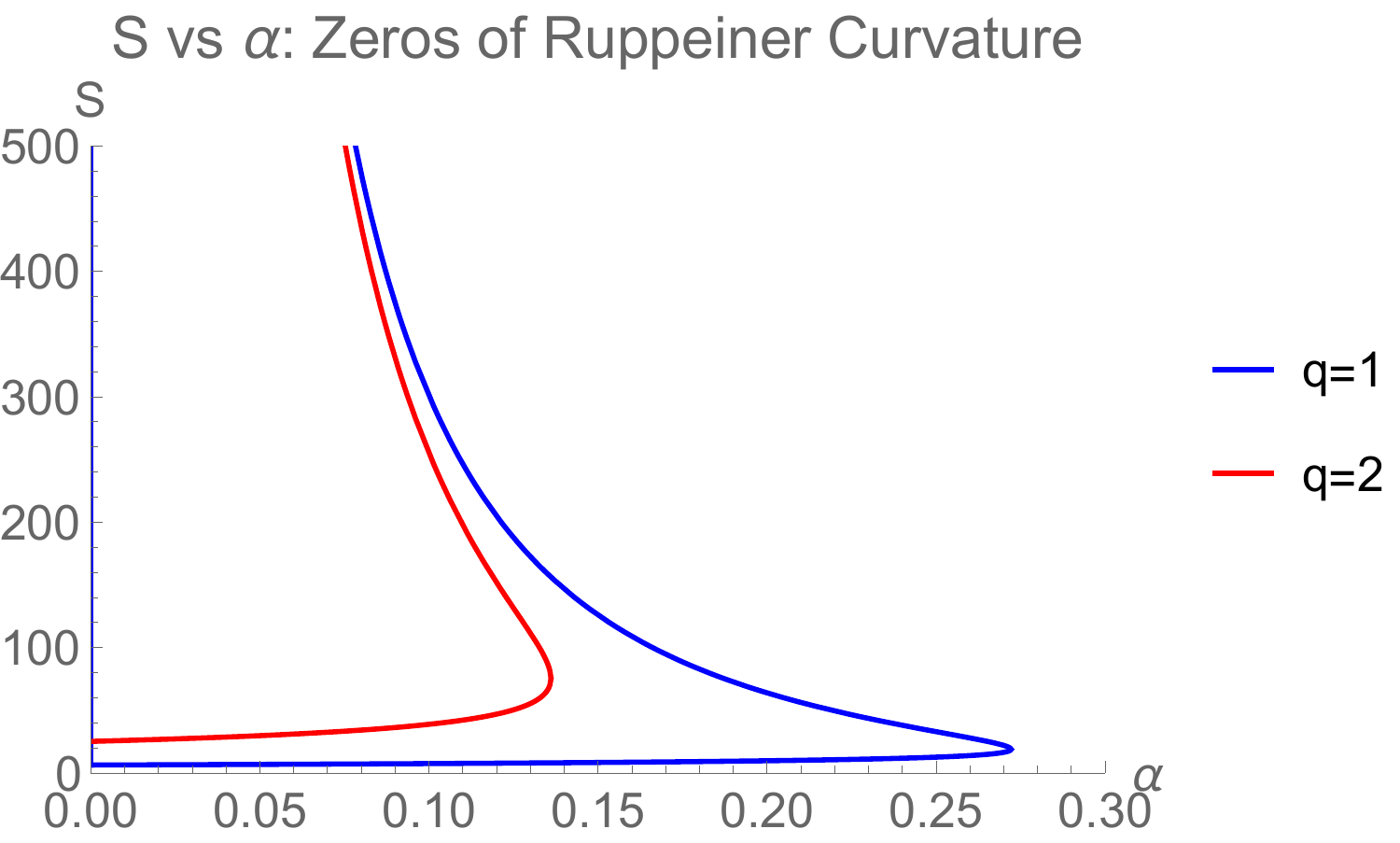}  		
		\caption{Points in the parameter space where the Ruppeiner Curvature is flat.}
        \label{s_vs_alpha}	
	\end{center}
%	\end{wrapfigure}
\end{figure}

Setting the rotation parameter in eqn (\ref{1}) to zero, we get a Reissner-Nordstr\"{o}m-AdS (RN-AdS) black hole surrounded by quintessence, whose lapse function $f(r)$ takes the form~\cite{Kiselev},
\begin{eqnarray}
&&    f(r) = 1 - \frac{m}{r} + \frac{q^2}{r^2} + \frac{r^2}{l^2} - \alpha r ,\\
&&  F=dA\,,\quad A=-\frac{q}{r} dt\, .
\end{eqnarray}  Making the usual identification that the ADM mass of the black hole and expressing it as function of entropy and pressure, one gets enthalpy to be,
\begin{equation}
H(S,P)=\frac{1}{2}\sqrt{\frac{S}{\pi}}\,\left(1+\frac{\pi q^2}{S}- \alpha \left(\sqrt{\frac{S}{\pi}}\right)+\frac{8 P S}{3}\right). \label{Hquint}
\end{equation} Note that in this non-rotating case, one has \(Q = q\). Thermodynamic volume can now be obtained from $V = (\partial H/\partial P)_{S,q,\alpha}$ as,
\begin{equation} \label{VS}
V = \frac{4 \pi r_+^3}{3}\, ,
\end{equation}
and is found to be same as geometric volume~\cite{Liu:2017baz}. With all definitions of fundamental thermodynamic variables available, the first law of black holes in the extended thermodynamic phase space can be written to be,
\begin{equation}
dM=dH=TdS+\Phi dq+VdP+\mathcal {A}d\alpha,\label{11}
\end{equation}%
where $\Phi = \left(\frac{\partial H}{\partial q}\right)_{S,P,\alpha}$ is the electric potential conjugate to charge $q$ and $\mathcal{A}=\left(\frac{\partial H}{\partial \alpha}\right)_{S,q,P}$ is a quantity conjugate to the parameter $\alpha$. It should be mentioned that the first law given in eqn (\ref{11}) is phenomenological, which makes the first law and Smarr relation consistent~\cite{Liu:2017baz}. A full derivation would require the use of field equations extending the techniques used in~\cite{Kastor} to the current situation involving quintessence. We will work in the canonical ensemble where \(dq = d\alpha = 0\). The Hawking temperature [eqn (\ref{11})] takes a simplified form,
\begin{equation}\label{T}
  T = \frac{1}{4\pi} \bigg( \frac{1}{r_+} - \frac{q^2}{r_+^3} - 2\alpha + 8\pi P r_+  \bigg).
\end{equation} Further, setting $J=0$ in eqn (\ref{eosqj}) and identifying the specific volume to be \(v = 2r_+\), one obtains the standard fluid-like equation of state \(P = P(v,T)\),
\begin{equation}\label{equationofstate}
  P = \frac{T}{v} + \frac{\alpha}{2\pi v} - \frac{1}{2\pi v^2} + \frac{2q^2}{\pi v^4},
\end{equation}
which corresponds to that of a non-ideal fluid, i.e. one with non-trivial interactions between molecules. Each of the last three terms in eqn (\ref{equationofstate}) signify interactions between the microstructures. The limit \(\alpha = 0\) gives the well known equation of state for standard RN-AdS black holes in \((3+1)\)-dimensions \cite{Kubiznak:2012wp}. Using the above equations for temperature and equation of state, the line elements in eqn (\ref{RuppeinerSP}) and (\ref{RuppeinerTV}) can be obtained. For instance, the \(2 \times 2\) metric tensor on the $(S,P)$-plane turns out to be,
\begin{equation}
g_{ij}= \left(
\begin{array}{cc}
 \frac{\sqrt{\pi } \left(3 \pi  q^2+S (8 P S-1)\right)}{2 S
   \left(\sqrt{\pi } \left(-\pi  q^2+8 P S^2+S\right)-2 S^{3/2} \alpha
   \right)} & \frac{8 \sqrt{\pi } S^2}{\sqrt{\pi } \left(-\pi  q^2+8 P
   S^2+S\right)-2 S^{3/2} \alpha } \\
 \frac{8 \sqrt{\pi } S^2}{\sqrt{\pi } \left(-\pi  q^2+8 P
   S^2+S\right)-2 S^{3/2} \alpha } & 0 \\
\end{array}
\right)
\end{equation}
where $i,j=S,P$. We can now give the analytical expressions of the Ruppeiner curvature for RN-AdS black holes surrounded by quintessence in the background, both on \((S,P)\) and \((T,V)\)-planes respectively as,
 \begin{equation} \label{RSP}
 R_{SP}=-\frac{\sqrt{\pi} (2 \pi q^2 - S) + S^{3/2} \alpha}{\sqrt{\pi} S (\pi q^2 - S (1 + 8 P S)) + 2 S^{5/2} \alpha} \, ,
 \end{equation}
 and
\begin{equation}\label{RTV}
 R_{TV}=\frac{8 \pi q^2 - 6^{2/3} \pi^{1/3} V^{2/3} + 3 V \alpha}{3\times6^{2/3} \pi^{4/3} T V^{5/3}} .
\end{equation}
 It should be noted that both $R_{SP}$ and $R_{TV}$ asymptotically diverge as the black hole becomes extremal, i.e. \(T=0\). The scalar $R_{SP}$ is plotted in figure-(\ref{RSplot}) (as a function of \(S\) for fixed \(P\). The scalar $R_{TV}$ shows identical behavior and hence is not shown here.
 %and (\ref{RTplot}) (as a function of \(V\) for fixed \(T\)), respectively.
On either of the thermodynamic planes, it is found that the Ruppeiner curvature admits zero crossings indicating  existence of points where the dominant kind of interactions can switch from attractive to repulsive and vice versa, since the sign of the Ruppeiner curvature indicates the nature of microscopic interactions. Let us first note that for \(\alpha = 0\), both $R_{SP}$ and $R_{TV}$ reduce to the previous expressions for the RN-AdS black hole obtained in \cite{Ghosh}. Furthermore, the curvature scalars given in eqns (\ref{RSP}) and (\ref{RTV}) can be checked to be equivalent so we may label them as \(R\). Second, for a given value of charge $q$, there is a possibility that for a value of $\alpha$, the Ruppeiner curvature changes sign twice. The points where $R=0$ occur at $S=13.8, 27.8$ for the value $\alpha=0.26, q=1$ in figure-(\ref{RSplot}). This is to be contrasted with the situation for RN AdS black holes with no quintessence~\cite{Wei2019a,Wei2019b,Ghosh}, where there is only one point where the Ruppeiner curvature crosses from repulsive to attractive type interactions of microstructures (plot for RN-AdS black holes is given as an inset in figure-(\ref{RSplot})). The condition \(R = 0\) occurs at points which are the physical solutions to the algebraic equation,
\begin{equation}\label{RNBound}
2 \pi ^{3/2} q^2+\alpha  S^{3/2}-\sqrt{\pi } S=0.
\end{equation}
The crossing points therefore depend on the parameters \(\alpha\) and \(q\) of the black hole. A plot of $S$ vs $\alpha$ shown in figure-(\ref{s_vs_alpha}) gives interesting insights.
As seen from figure-(\ref{s_vs_alpha}) the number of zero crossing points are either zero, one or two depending for a particular case on how many real solutions eqn (\ref{RNBound}) admits based on the values of \(\alpha\) and \(q\). As mentioned above, the existence of one zero of the Ruppeiner curvature at small value of $S$\footnote{Indicating that the smaller black holes are dominated by repulsive interactions among microstructures \cite{Wei2019a,Wei2019b}.} at which the attractive and repulsive interactions balance each other is well known for the RN-AdS system~\cite{Ghosh}. The cases where there's a second zero (see figure-(\ref{RSplot})) at a higher entropy (or equivalently horizon size), are new and of particular interest here, because they are associated with the presence of quintessence in the background. In fact, there are two further remarkable results. First is the existence of a point on the $S$-axis (the red curve in figure-(\ref{RSplot})) corresponding to a threshold value of $\alpha=\alpha_0$ (depending on $q$), where the $R$ starts out positive, barely vanishes at a value of $S$ and continues to be positive after that. This threshold value occurs at the following point,
\begin{equation} \label{threshold}
\alpha_0 = \frac{0.27216}{q}, \quad S = 6 \pi \, q^2 \, .
\end{equation}
The existence of such an $\alpha_0$ can be seen from figure-(\ref{s_vs_alpha}) as well (the extreme point at which only one solution for $S$ exists, for a given value of $\alpha$). Second, for any value beyond $\alpha_0$, $R$ is always positive, indicating the domination of repulsive interactions among microstructures, which overcome the attractive interactions intrinsic to the larger RN-AdS black holes. It will be seen that such a crossing can be explained by considering long range repulsive interactions to be associated with quintessence.

\smallskip

\begin{figure}[h]
%	 \begin{wrapfigure}{l}{0.3\textwidth}
	\begin{center}
		\centering
		\includegraphics[width=4.6in]{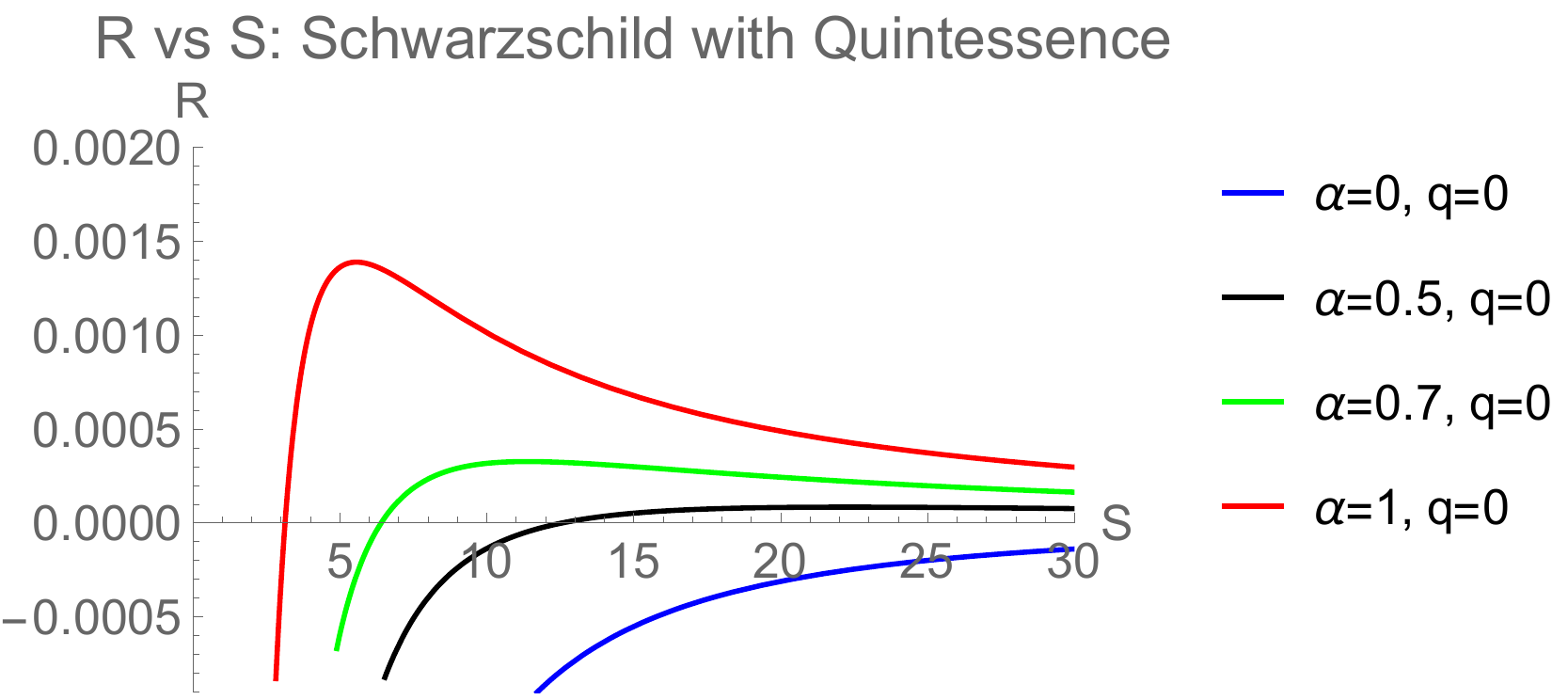}  		
		\caption{Ruppeiner curvature for Schwarzschild-AdS black hole surrounded by quintessence on the \((S,P)\)-plane as a function of \(S\) for fixed \(P\).}
        \label{SAdSplot}	
	\end{center}
%	\end{wrapfigure}
\end{figure}

We now consider the \(q=0\) limit where the system is a Schwarzschild-AdS black hole surrounded by quintessence. As seen from the $q=0$ limit of the Ruppeiner curvature in eqn (\ref{RSP}), quite remarkably, quintessence introduces a repulsion dominated region in the otherwise purely attractive Schwarzschild-AdS black hole. In this case one gets the crossing point of the Ruppeiner curvature to be simply at,
\begin{equation}\label{SAdS}
  r_+ = \frac{1}{\alpha} ,
\end{equation} or equivalently \(\rho_0 = \alpha/2\) in terms of the density (defined as reciprocal of specific volume). Note that the crossing point increases with decrease in quintessence \(\alpha\) signifying the repulsive nature of the microstructures introduced due to quintessence. The Ruppeiner curvature is plotted as a function of \(r_+\) in figure-(\ref{SAdSplot}).
There is another important novel consequence of the quintessence parameter $\alpha$ on the thermodynamic volume of Schwarzschild AdS black hole. In general in fluids, such as the vdW fluid, there is a minimum volume dictated by the fact that temperature does not become negative. This can be obtained from the equation of state evaluated at $T=0$. Although Schwarzschild-AdS black holes do not show vdW behavior, in the presence of quintessence, there is a possibility of minimum volume given by,
\begin{equation}
V_{min}=\frac{6 \alpha ^3\pm \sqrt{\alpha ^2 \left(36 \pi  P-6 \alpha
   ^2\right)^2-1152 \pi ^3 P^3}-36 \pi  \alpha  P}{576 \pi ^2\,
   P^3} .
\end{equation}
It is clear that there is no minimum volume when $\alpha=0$, corresponding the existence of only attractive interactions in the pure Schwarzschild-AdS (without quintessence) case \cite{Xu}.

\subsection{Slowly rotating black holes in AdS}\label{slowly}
\begin{figure}[h]
%	 \begin{wrapfigure}{l}{0.3\textwidth}
%	\begin{center}
		\centering
		\includegraphics[width=4.6in]{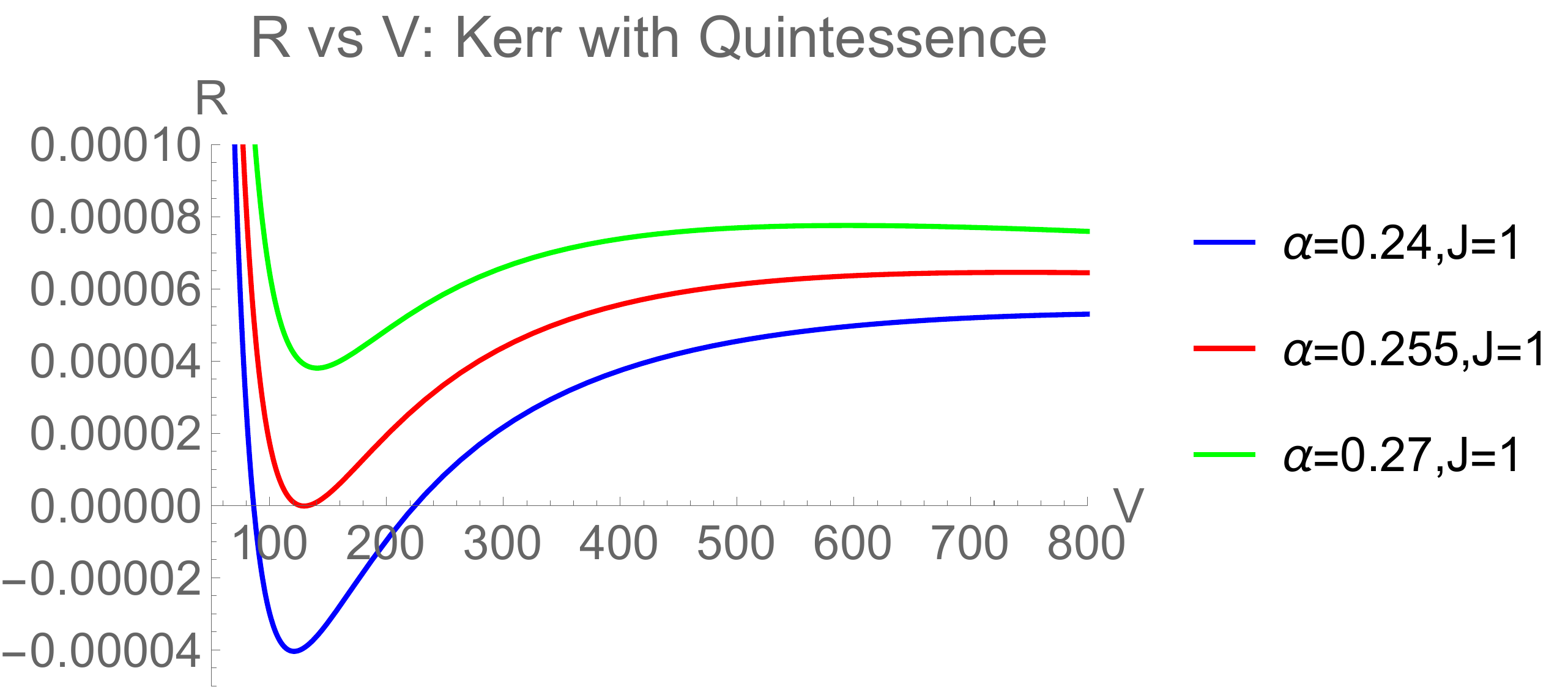}  		
\caption{Ruppeiner curvature for Kerr-AdS black hole surrounded by quintessence on the
		\((T,V)\)-plane as a function of \(V\) for fixed \(T\) and angular momentum $J$.}
        \label{KAdSplot}	
%	\end{center}
%	\end{wrapfigure}
\end{figure}

The general expression for Ruppeiner curvature can be obtained analytically in the case $Q=0, J \neq 0$, but complicated and not shown here. Instead, we plot the result directly, as shown in figure-(\ref{KAdSplot}). We find that behavior of $R$ is qualitatively similar to the case of charged black holes in AdS, with $J$ playing the role of $Q$. Therefore, even in present case, the presence of the quintessence parameter ensures that the Ruppeiner scalar has two zero crossings and hence one expects this to be a generic feature in other black hole systems too, including the case of a general rotating black hole.
Furthermore, there is a minimum volume dictated by the equation of state given in eqn (\ref{eosqj}) so that the temperature is not negative, which cannot be obtained analytically, but one can obtain its values numerically. For instance, considering the case $\alpha=0.24$, the minimum volume is $V_{min}=3.08$. Keeping this in mind, even if there are additional points where $R$ goes to zero below this volume, they are ignored as they are not physical from the point of view of the equation state.

\section{Microstructures of black holes and a mean field description} \label{mean}
%%%%%%%%%%%%%%%%%%%%%%%%%%%%%%%%%%%%%%%%%%%%
Analysis of Ruppeiner curvature of AdS black holes  in last section, shows a subtle interplay between the attraction and repulsion dominated regimes, which heuristically suggests the presence of both attractive and repulsive interactions in the system, akin to the behavior of microstructures in a vdW fluid~\cite{Wei2019a,Wei2019b}. Despite long standing efforts, the true nature of microscopic degrees of freedom of black holes are not known yet. Following Boltzmann's lines, ``if you can heat it, it has microstructure'', current studies have tried to obtain whatever valuable information from connection between thermodynamic and gravitation properties of black holes. The black hole chemistry paradigm~\cite{K,Kubiznak:2016qmn} has shown that charged black holes in AdS have a rich phase structure similar to vdW systems, suggesting a possible molecular kind of microstructure emerging from the degrees of freedom for black holes as well. However, the analysis of thermodynamic curvature shows that the behavior of microstructures of the two systems is not the same, with the case of black holes being more subtle~\cite{Wei2019a,Wei2019b,AR,Ghosh}. There are also proposals that a general interacting system can probably also be modeled as a binary mixture \cite{AR,Ghosh} of both attractive and repulsive microstructures which share the degrees of freedom of the total entropy. Here, we take these issues forward and show that within a mean field approximation, the interactions can be described by an effective mean field interaction potential whose extremum points correspond to points where the Ruppeiner metric is flat. However, we particularly emphasize on the fact that any analysis in this spirit is phenomenological and is possible due to the resemblance of the thermodynamic structure of AdS black holes with that of hydrostatic systems such as the vdW fluid. Therefore, we shall be exploring a fluid-like description of black hole microstructures which may be thought of as being analogous to the molecules present in a fluid. Since \(v=2r_+\) corresponds to the specific (fluid) volume of the black hole, a number density \(\rho = 1/v\) can be naturally associated with the system~\cite{interactions,Wei:2015iwa,Bairagya:2019lxq}. One can then alternatively write the equation of state [eqn (\ref{equationofstate})] in terms of this density as,
\begin{equation}\label{equationofstaterho}
  P = \rho T + \bigg(\frac{\alpha}{2\pi}\bigg)\rho - \bigg(\frac{1}{2\pi}\bigg)\rho^2 + \bigg(\frac{2q^2}{\pi}\bigg) \rho^4.
\end{equation}
The first term, which depends on the temperature is clearly the kinetic energy density\footnote{Indeed from the equipartition theorem, \(E \sim T\) and one can see that the equation of state for the ideal gas, \(P = \rho T\) equivalently implies \(P \sim \rho E\) making it a kinetic energy density.} whereas the remaining terms signify non-trivial microscopic interactions. In fact, one can think of the remaining terms to correspond to a mean field potential energy density containing coarse grained information of the interactions and is given by,
\begin{equation}\label{u}
  u(\rho) = A\rho - B\rho^2 + C \rho^4,
\end{equation} where \(A = \alpha/2\pi\), \(B=1/2\pi\) and \(C = 2q^2/ \pi \) are all positive constants. It is then expected that the extremum of such a mean field interaction potential would dictate the points of no effective interactions. This is indeed true and eqn (\ref{RNBound}) whose physical solutions correspond to the crossing points of the Ruppeiner curvature can alternatively be obtained as,
\begin{equation}\label{dP}
  \bigg(\frac{\partial u}{\partial \rho}\bigg) = 0 .
\end{equation}
In the following subsection, we demonstrate the utility of the mean field interaction potential approach, by considering the special case \(\alpha=0\), i.e. RN-AdS black holes in the absence of quintessence. The cases with non-trivial dependence on quintessence are discussed in the next subsection.

%%%%%%%%%%%%%%%%%%%%%%%%%%%%%%%%%%%%%%%%%%%%
\subsection{RN-AdS black holes} \label{meanRN}
\begin{figure}[h]
%	 \begin{wrapfigure}{l}{0.3\textwidth}
	\begin{center}
		\centering
		\includegraphics[width=4.6in]{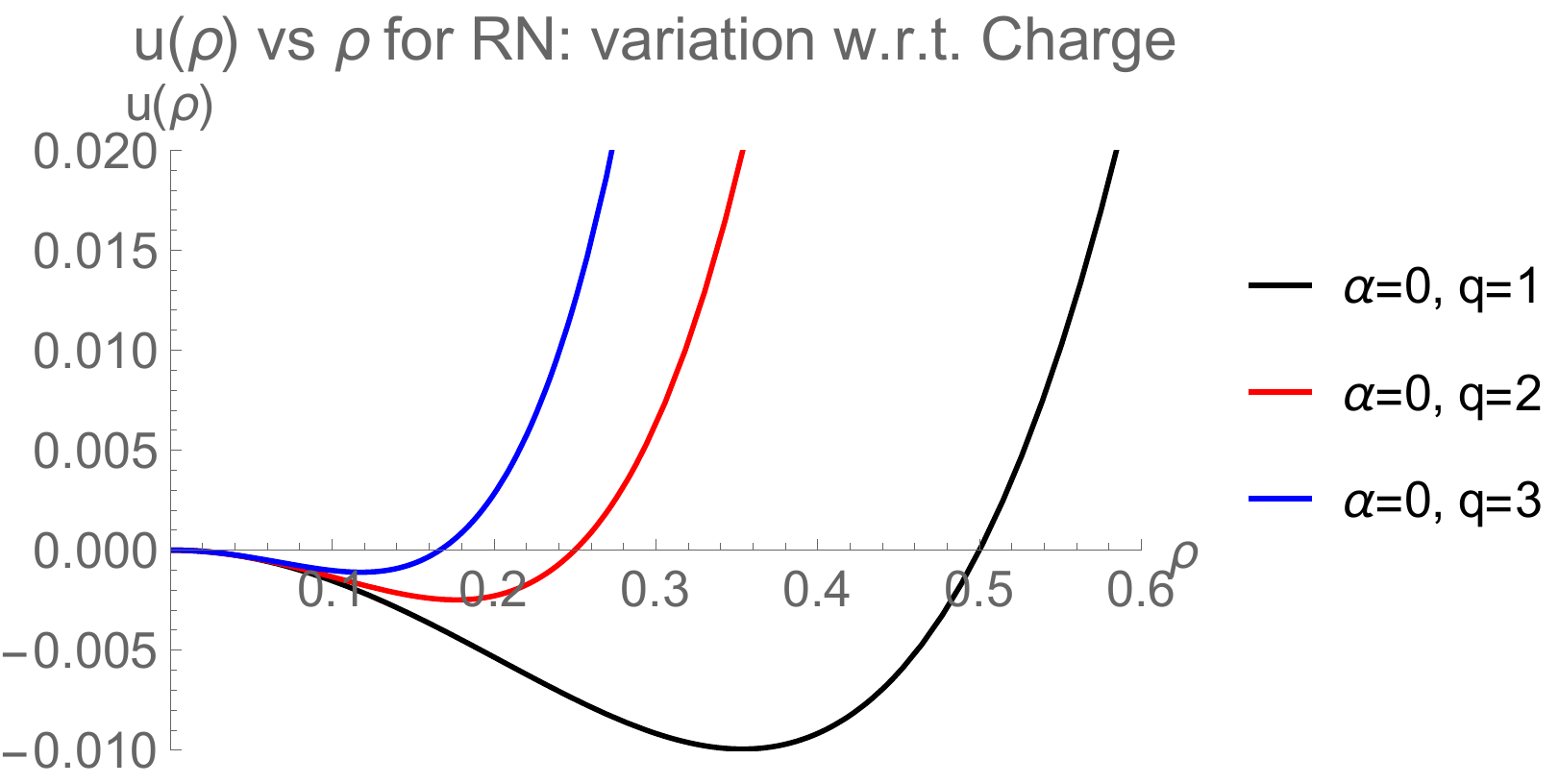}  		
		\caption{Mean field interaction potential \(u(\rho)\) as a function of \(\rho\) for various values of electric charge \(q\) with \(\alpha=0\).}
        \label{u_vs_rho_rn_charge}	
	\end{center}
%	\end{wrapfigure}
\end{figure}

%%%%%%%%%%%%%%%%%%%%%%%%%%%%%%%%%%%%%%%%%%%%
For the case of RN-AdS black holes without quintessence in the background, the mean field interaction potential is given as,
\begin{equation}\label{RNu}
  u(\rho) = -\bigg(\frac{1}{2\pi}\bigg)\rho^2 + \bigg(\frac{2q^2}{\pi}\bigg) \rho^4,
\end{equation} and is plotted as a function of \(\rho\) for various values of the electric charge \(q\) in figure-(\ref{u_vs_rho_rn_charge}). In all the cases, there exists a minimum\footnote{Since \(\rho=0\) (or equivalently \(v = \infty\)) is not physical, the only physical point at which this mean field interaction potential admits an extremum is the minimum.} of the potential where the attractive and repulsive interactions balance out. This point corresponds to \(\rho_0 = \sqrt{1/8q^2}\) reproducing the previously known result for the horizon radius, \(r_+ = \sqrt{2}|q|\) \cite{Ghosh}. The first derivative of \(u(\rho)\) is positive for \(\rho>\rho_0\) whereas it is negative for \(\rho<\rho_0\), translating respectively to repulsion and attraction dominance. This goes well with previously known conclusions for RN-AdS black holes where there exists a long range attraction and a short range repulsion \cite{Wei2019a}: \(\rho_0\) represents the density at which the repulsion and attraction balance out whereas compressing the fluid to a smaller volume and hence, greater density leads to dominance of repulsion and vice versa. Moreover, since for a fluid\footnote{The fluid in our discussions always assumes three spatial dimensions, irrespective of the number of spacetime dimensions in which the black hole (whose thermodynamics gets mapped into that of the fluid) resides in.}, the density scales as \(\rho \sim r^{-3}\) where \(r\) is a length scale typically of order of the mean free path of the molecules\footnote{With the assumption that the fluid density is spatially uniform, without any loss of generality one can consider some arbitrarily small length scale \(r\) of the order of the mean free path of the molecules such that in a volume \(v \sim r^3\), the number of fluid molecules is constant.}, the form of the mean field interaction potential [eqn (\ref{RNu})] suggests towards an intermolecular interaction of the Lennard-Jones type\footnote{See~\cite{RupMay} for a deeper connection between zero crossings of Ruppeiner curvature and intermolecular interactions involving Lennard-Jones type potentials in a different context.} as pointed out in \cite{LennardJonnes,Wei2019b},
\begin{equation}
  V(r) = -\frac{c}{r^6} + \frac{d}{r^{12}},
\end{equation} where \(c\) and \(d\) are appropriate positive constants. The mean field approach is therefore consistent with the previously known results for RN-AdS black holes and provides a satisfactory picture of interactions among microstructures.

%%%%%%%%%%%%%%%%%%%%%%%%%%%%%%%%%%%
\subsection{Effect of dark energy on the mean field interaction potential}\label{quint}
%%%%%%%%%%%%%%%%%%%%%%%%%%%%%%%%%%%

We shall now consider the effect of quintessence on the microstructures of AdS black holes. The mean field interaction potential in this case takes the form,
\begin{equation}\label{Salphau}
  u(\rho) = \bigg(\frac{\alpha}{2\pi}\bigg)\rho - \bigg(\frac{1}{2\pi}\bigg)\rho^2.
\end{equation}\begin{figure}[h]
%	 \begin{wrapfigure}{l}{0.3\textwidth}
	\begin{center}
		\centering
		\includegraphics[width=4.6in]{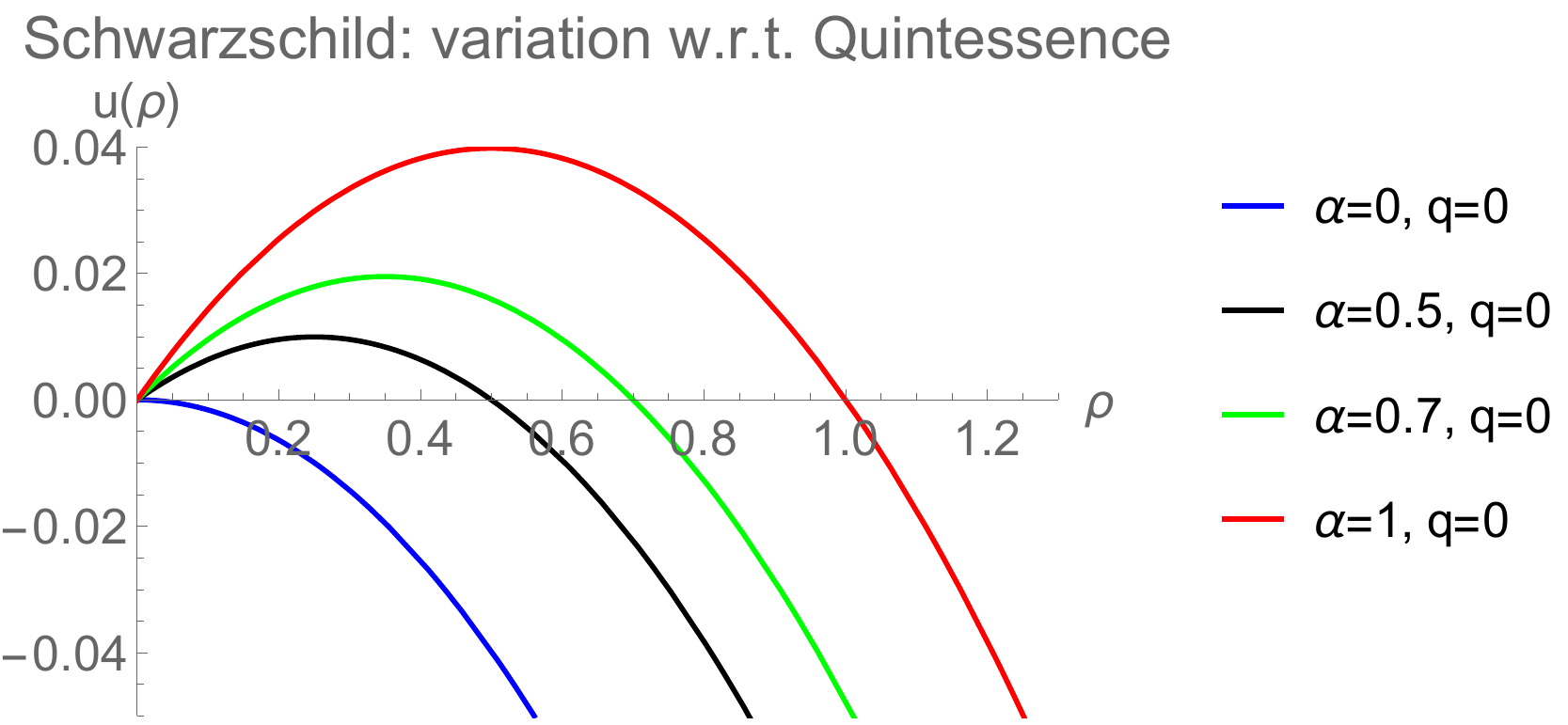}  		
		\caption{Mean field interaction potential \(u(\rho)\) as a function of \(\rho\) for various values of \(\alpha\) with \(q=0\).}
        \label{u_vs_rho_sads_alpha}	
	\end{center}
%	\end{wrapfigure}
\end{figure}Interestingly, the crossing point of the Ruppeiner curvature corresponds to a maximum, i.e. an unstable equilibrium of the mean field interaction potential,
\begin{equation}
  \frac{\partial^2 u}{\partial \rho^2} = -\frac{1}{\pi}.
\end{equation}
A plot of the mean field interaction potential is shown in figure-(\ref{u_vs_rho_sads_alpha}). This means that with \(\rho_0\) being the value of the density at which attraction and repulsion balance out each other, a higher density would mean attraction dominance while a lower density would imply the dominance of repulsion: a conclusion which is completely opposite from that from the pure RN-AdS case discussed in the previous subsection. A fluid such as this one would admit a microscopic intermolecular potential whose repulsive part has a range which is longer than that of the attractive part. By our arguments from the previous subsection, it is suggestive from the form of eqn (\ref{Salphau}) that such a fluid would have a microscopic intermolecular potential of the form,
\begin{equation}
  V(r) = \frac{b}{r^3} - \frac{c}{r^6},
\end{equation} where \(b\) and \(c\) are both positive constants. The first and second terms respectively correspond to the repulsive and attractive parts of the intermolecular interaction. Naively, such an observation looks bizarre, but can be explained from the fact that quintessence, which corresponds to a negative pressure responsible for the expansion of the universe is essentially a long ranged repulsive interaction with a range that is naturally longer than the range of attractive interactions between microstructures of the Schwarzschild-AdS black hole. Microstructures associated with quintessence are therefore not just repulsive, but have long ranged interactions, which typically go as an inverse cube of the distance.

\smallskip

\begin{figure}[h]
%	 \begin{wrapfigure}{l}{0.3\textwidth}
	\begin{center}
		\centering
		\includegraphics[width=4.6in]{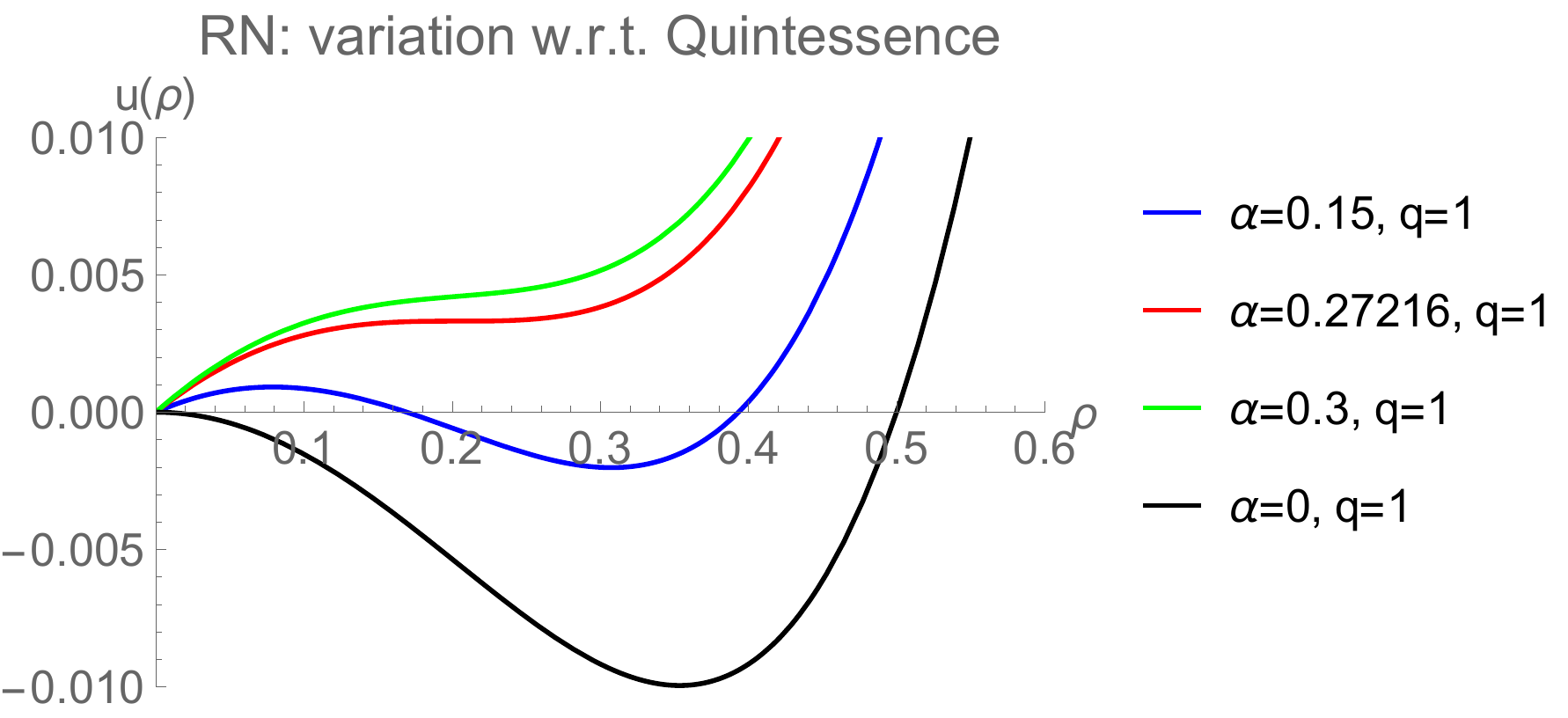}  		
		\caption{Mean field interaction potential \(u(\rho)\) as a function of \(\rho\) for various values of \(\alpha\) with \(q=1\).}
        \label{u_vs_rho_rn_alpha}	
	\end{center}
%	\end{wrapfigure}
\end{figure}

The presence of a non-trivial electric charge makes the behaviour of the mean field interaction potential all the more interesting and is shown in figure-(\ref{u_vs_rho_rn_alpha}).
  Note that the black curve corresponds to the RN-AdS case without quintessence, i.e. \(\alpha=0\). The red one corresponds to \(\alpha=\alpha_0=0.27216\) (with \(q=1\)) and one can see that for \(\alpha > \alpha_0\) (the green curve), there are no extremum points of \(u(\rho)\) indicating absence of zero crossings of the Ruppeiner curvature. In fact, for this case, the first derivative of the mean field interaction potential is positive definite. Such a situation dictates complete absence of attraction dominance from the system and the Ruppeiner curvature is wholly positive. The case with \(\alpha < \alpha_0\) (the blue curve) is the most interesting one where there are two extremum points now as compared to the previous cases with zero crossings where there was either a maximum or a minimum. This means that there are now two equilibrium points where the net interactions balance out. Let us say these points correspond to the values \(\rho_{01}\) and \(\rho_{02}\) with \(\rho_{01} < \rho_{02}\). At \(\rho_{01}\), the equilibrium is an unstable one, meaning that further compression of the fluid would lead to attraction rather than repulsion and vice versa. It is around this unstable equilibrium that the third term of eqn (\ref{u}) has little consequence\footnote{The presence of electric charge would of course shift the position of this maximum but does not change the qualitative behavior of \(u(\rho)\) around the maximum.} and equilibrium is reached primarily as balance of interactions which the first and second terms of eqn (\ref{u}) signify. This is similar to the case of the neutral Schwarzschild-AdS black hole surrounded by quintessence. The other equilibrium point, which exists only if we consider non-trivial charge is a minimum at \(\rho_{02}\) and is hence stable. This equilibrium is reached primarily from a balance of interactions due to the second and third terms in eqn (\ref{u}) while the first term has very little to do with the behavior of this minimum.

\smallskip

In this paper, we restricted to the special case where the quintessence equation of state parameter took the value $\omega_q = -2/3$. If instead, one would choose other values of \(\omega_q\), the quintessence dependent term appearing in the mean field interaction potential would get altered in the power of \(\rho\) (which is linear in the present case, see eqn (\ref{u})) changing the overall shape of \(u(\rho)\) and hence the location of the zeros of the Ruppeiner curvature discussed in this work. Such cases might be interesting to pursue in the future, although for the form of geometry in eqn (\ref{1}) and equation of state considered in this paper, it would imply that it is always associated with short ranged repulsive interactions. Therefore, apart from changes in the locations at which the potential shows extremum, the general conclusions obtained from the present work should remain unaltered even if calculations are done with other values of \(\omega_q\) in that specified range.

\section{Remarks}\label{remarks}
%%%%%%%%%%%%%%%%%%%%%%%%%%
In this work, we have focused on probing the nature of interactions among the microstructures for asymptotically AdS black holes surrounded by quintessence using methods of thermodynamic geometry. From the empirical behavior of thermodynamic scalar $R$ charged black holes in \(d \geq 4\) are associated with both attraction and repulsion dominated regions ~\cite{Wei2019a,Wei2019b,Ghosh}. In this paper we considered various limiting case of the equation of state in~\cite{Sharif:2020hid} and studied the effect of dark energy on thermodynamic geometry and microstructures of neutral, charged and slowly rotating black holes. It may be possible to phenomenologically model the black hole microstructures as composed of two distinct kindse from a fluid perspective, where there is an exact analogy of black hole system with vdW system\cite{Kubiznak:2012wp,AR,Wei2019b,Ghosh}. It is known from earlier studies that the repulsive interactions are rather short ranged as compared to their attractive counterparts, such that the overall microscopic interaction potential is suggestively of the Lennard-Jonnes type \cite{LennardJonnes,Wei2019b}. Following on these ideas, we developed an effective mean field potential approach for a generic black hole system in AdS in section-(\ref{mean}), such that the details of regions of domination of interactions among microstructures correspond to the equilibrium points of this potential. We also learn that quintessence leads to long ranged repulsive effects. In extended thermodynamics, where the fluid analogy of the black hole system is well established~\cite{Kubiznak:2012wp,45,Li}, we see that the repulsive interaction due to quintessence is analogous to an inverse cube microscopic interaction among the fluid molecules,
\begin{equation}
  V_{quint}(r) \sim \frac{1}{r^3}.
\end{equation}
It would be interesting to see the effects of quintessence on the microstructures for black holes in higher derivative theories such as Gauss-Bonnet-AdS black holes or even the more general Lovelock-AdS and pure Lovelock-AdS black holes, leading to a complete classification of attraction and repulsion dominated regions for wide variety of black holes. Another important avenue for future work is to include thermal fluctuations, as it is known that fluctuations give additional contributions and modify the points where the Ruppeiner scalar $R$ goes to zero. The effect of statistical fluctuations on thermodynamics and phase transitions is well studied and is known to change the definitions of entropy and free energy, with particularly novel consequences for large black holes~\cite{18,18a,21,21a,22a,23a,Sharif:2020hid}. Thus, it should be intriguing to consider additional effects due to thermal fluctuations on the thermodynamic geometry of black holes with quintessence studied in this work.

\section*{Acknowledgements}
A.S. wishes to thank the Council of Scientific and Industrial Research (CSIR), Government of India, for financial support. A.G. would like to acknowledge the financial support received from IIT Bhubaneswar in the form of an institute research fellowship. The authors are grateful to the anonymous referees for a careful reading of the manuscript and their useful comments which have led to an improvement of the article.

\appendix

\section*{Appendix: Metric structures in the extended thermodynamic phase space}\label{geometry}
%%%%%%%%%%%%%%%
It is well appreciated that the thermodynamic phase space assumes the structure of a contact manifold (see for example \cite{Mrugala,Bravetti,contactBH}). Recall that a contact manifold is the pair \((\mathcal{M},\eta)\) where \(\mathcal{M}\) is a smooth manifold of \((2n+1)\)-dimensions and \(\eta\) is a one form such that its kernel is a distribution of hyperplanes, i.e. co-dimension one which is completely non-integrable in the Frobenius sense. This means that the following condition holds,
\begin{equation}\label{contactform}
  \eta \wedge (d\eta)^n \neq 0.
\end{equation}
The one form \(\eta\) is known as the contact form whose kernel which is the completely non-integrable hyperplane distribution called the contact structure. Any odd dimensional smooth manifold with a contact structure admits a one form \(\eta\) satisfying eqn (\ref{contactform}) and is called a contact manifold. Associated with \(\eta\), there exists a unique and global vector field \(\xi\) known as the Reeb vector field whose flow preserves the contact structure. It is defined uniquely as,
\begin{equation}\label{xidef}
  \eta(\xi) = 1, \hspace{3mm} d\eta(\xi,.)=0.
\end{equation}
Analogous to the symplectic case, there exists a Darboux's theorem for the case of contact manifolds which states that, any \((2n+1)\)-dimensional contact manifold \((\mathcal{M},\eta)\) is equivalent to \((\Re^{2n+1},\eta)\) with,
\begin{equation}\label{etalocal}
  \eta = ds - p_i dq^i, \hspace{3mm} i \in \{1,2,...,n\},
\end{equation} where the coordinates \((s,q^i,p_i)\) are known as Darboux coordinates. In other words, on any contact manifold \((\mathcal{M},\eta)\), one can define locally Darboux coordinates such that the contact form is given by eqn (\ref{etalocal}). It is easy to verify that in these local coordinates, the Reeb vector field \(\xi\) is given as,
\begin{equation}
  \xi = \frac{\partial}{\partial s}.
\end{equation}
Furthermore, there are a very special class of submanifolds of a contact manifold \((\mathcal{M},\eta)\) which are of interest especially from the thermodynamics perspective. They are the integral submanifolds of maximum dimension such that \(\eta=0\) when restricted to the submanifold. In other words, if \(q^i\) and \(p_i\) are to be treated as conjugate variables, it is easy to see from eqn (\ref{etalocal}) that such a submanifold cannot contain a conjugate pair and hence correspond to the familiar notion of configuration spaces from mechanics. For a particular Legendre submanifold \(L\) having coordinates \((q^i,p_j)\) where \(i \in I, j \in J\) with \(I\) and \(J\) being a disjoint partition of the index set \(\{1,2,....,n\}\), the local structure is always given as,
\begin{equation}\label{Llocal}
  p_i = \frac{\partial F}{\partial q^i}, \hspace{3mm} q^j = - \frac{\partial F}{\partial p_j}, \hspace{3mm} s = F - p_j \frac{\partial F}{\partial p_j}.
\end{equation} In this context \(F=F(q^i,p_j)\) is known as the generator of \(L\) and it is therefore clear that all Legendre submanifolds are \(n\)-dimensional. It was shown long back~\cite{Sasaki,Hatakeyama,SasakiHatakeyama} that any contact manifold can be associated with a Riemannian metric structure which satisfies some compatibility conditions with the contact form. The metric is a bilinear, symmetric as well as non-degenerate structure. It can be verified that the generic choice, \(G = \eta - dq^i dp^i\) satisfies all these three basic requirements and since for an arbitrary Legendre submanifold \(L\), one has \(\eta_L=0\) by definition, restricting \(G\) to \(L\) gives the local expression from eqns (\ref{Llocal}),
\begin{equation}\label{GL}
  G|_L = -dp_i dq^i|_L = \frac{\partial^2 F}{\partial p_j \partial p_j'} dp_j dp_j' - \frac{\partial^2 F}{\partial q^j \partial q^{j'}} dq^j dq^{j'}.
\end{equation}
The metric on a Legendre submanifold \(L\) is therefore defined from the Hessian of the generator of \(L\). There are of course other ways of defining a symmetric, bilinear and non-degenerate metric structure in a contact manifold which would not be of interest to us.\\

%%%%%%
\subsection*{Connection with thermodynamics and derivation of the line elements of the Ruppeiner metric} \label{sN2metric}
%%%%%%%
%%%%
%Having recalled some basic results from contact geometry required for our purposes, let us make a connection with thermodynamics of black holes in $(s,N^2)$-plane.
Recall the first law given in eqn (\ref{FLbulk}),
\begin{equation}\label{firstlawcontact}
  dS - \beta dH + \beta V dP = 0,
\end{equation}
re-written to cast entropy as the fundamental thermodynamic potential with \(\beta = 1/T\). A comparison of eqn (\ref{firstlawcontact}) with the condition \(\eta_L = [ds - p_i dq^i]_L = 0\), which defines a Legendre submanifold \(L\) of a contact manifold \(\mathcal{M}\), leads to the straightforward identification that thermodynamic variables can be regarded as local coordinates on a contact manifold. In other words, the thermodynamic phase space in this case is a five dimensional contact manifold with a contact form, \(\eta = dS - \beta dH + \beta V dP\) defined on it with the identifications, \(s = S, p_1 = \beta, q^1 = H, p_2 = - \beta V, q^2 = P\). In equilibrium, eqn (\ref{firstlawcontact}) holds exactly, meaning that the physical system lies on a Legendre submanifold whose local structure is the following (by comparison with eqn (\ref{Llocal})),
\begin{equation}\label{Llocal1}
  \beta = \frac{\partial S}{\partial H}, \hspace{3mm} \beta V = - \frac{\partial S}{\partial P}.
\end{equation} Clearly, \((H, P)\) are the independent coordinates on this submanifold whereas, \(S=S(H,P)\) is the appropriate generating function. Points on this submanifold correspond to different thermodynamic equilibrium states of the black hole. The Hessian metric whose generic form is given in eqn (\ref{GL}), in this case gives,
\begin{equation}\label{R11111}
  dl_R^2 = - \frac{\partial^2 S}{\partial q^i \partial q^{i'}} dq^i dq^{i'},
\end{equation} where \(i,i' = 1,2\) and \(q^1 =H, q^2 = P\) are known as fluctuation coordinates. The form of the metric in eqn (\ref{R11111}) is known as the Ruppeiner metric in literature. It should therefore be clear that the Ruppeiner metric is one special case of more general Hessian metrics [eqn (\ref{GL})] that one can define on the thermodynamic phase space. What this means is that, we could have written our first law equally well as, \(dH - Tds - V dP = 0\) wherein the resulting metric would turn out to be a negative Hessian of the enthalpy \(H\) with fluctuation coordinates being \(q^1 = S, q^2 = P\). The curvature scalar of the Ruppeiner metric holds a special significance because its sign (and the points at which it diverges) can provide useful information about the thermodynamics of a system of interest. However, the scalar curvature associated with an arbitrary Hessian metric [eqn (\ref{GL})] may not have such nice properties. Therefore, for our analysis, we would specifically focus on the Ruppeiner metric and analyze its associated scalar curvature.

\smallskip

We now show how the Ruppeiner metrics presented earlier in eqns (\ref{RuppeinerSP}) and (\ref{RuppeinerTV}) can be obtained, with the starting point as eqn (\ref{R11111}). Following this, the connection of these line elements when other thermodynamic potentials are taken as the starting  points is also indicated. Let us start by trying to re-write eqn (\ref{R11111}) such that for calculational purposes, one can choose either \(S\) and \(P\) or \(T\) and \(V\) to be independent. Note that from eqn (\ref{firstlawcontact}), we have the identifications \((q^1,p_1,q^2,p_2) = (H, \beta, -\beta V, P)\) so that,
\begin{equation}
  dq^1 = dH, \hspace{3mm} dp_1 = d\beta, \hspace{3mm} dq^2 = -\beta dV - V d\beta, \hspace{3mm} dp_2 = dP,
\end{equation} where the Ruppeiner line element is given by eqn (\ref{R11111}). Since on the Legendre submanifold \(L\) representing the black hole, the line element is given by \(dl_R^2 = -dp_i dq^i|_L = -dp_1 dq^1 - dp_2 dq^2\) we have,
\begin{equation}
  dl_R^2 = -d\beta dH + (\beta dV + V d\beta) dP,
\end{equation}
which gets simplified to,
\begin{equation}
  dl_R^2 = (-dH + V dP) d\beta + \beta dV dP.
\end{equation}
Now substituting eqn (\ref{firstlawcontact}) in the first term above, the entire expression simplifies further to,
\begin{equation} \label{lineelementgeneric1}
  dl_R^2 = - \frac{1}{\beta} dS d\beta + \beta dV dP = \frac{1}{T}(dS dT + dV dP).
\end{equation}
At this stage, we choose \(S\) and \(P\) to be independent, i.e.
\begin{equation}
  T = T(S,P), \hspace{3mm} V = V(S,P),
\end{equation} and consequently,
\begin{equation}
  dT = \bigg(\frac{\partial T}{\partial S}\bigg)_{P}dS + \bigg(\frac{\partial T}{\partial P}\bigg)_S dP, \hspace{3mm} dV = \bigg(\frac{\partial V}{\partial S}\bigg)_{P}dS + \bigg(\frac{\partial V}{\partial P}\bigg)_S dP,
\end{equation} which upon substitution into eqn (\ref{lineelementgeneric1}) gives the final form identical to eqn (\ref{RuppeinerSP}) which can also be compactly written down as,
\begin{equation}
  dl_R^2 = \frac{1}{T}\frac{\partial^2 H}{\partial y^i \partial y^j}dy^i dy^j,
\end{equation} where \(y^1 = S\) and \(y^2 = P\). This shows that eqn (\ref{Weinhold}) is conformally related to the Ruppeiner metric [eqn (\ref{Ruppeiner})]. Similarly picking up \(T\) and \(V\) to be independent, one can write,
\begin{equation}
  S=S(T,V), \hspace{3mm} P=P(T,V).
\end{equation} This means,
\begin{equation}
  dS = \bigg(\frac{\partial S}{\partial T}\bigg)_{V}dT + \bigg(\frac{\partial S}{\partial V}\bigg)_T dV, \hspace{3mm}dP = \bigg(\frac{\partial P}{\partial T}\bigg)_{V}dT + \bigg(\frac{\partial P}{\partial V}\bigg)_T dV,
\end{equation} which upon substitution into eqn (\ref{lineelementgeneric1}) yields the line element given in eqn (\ref{RuppeinerTV}) or more compactly,
\begin{equation}
  dl_R^2 = \frac{1}{T}\frac{\partial^2 F}{\partial y^i \partial y^j}dy^i dy^j,
\end{equation} where \(F=F(T,V)\) is the Helmholtz free energy and one has \(y^1 = T, y^2 = V\).

\end{document}